\documentclass[a4paper,12pt]{scrartcl}%
\usepackage{amsfonts}
\usepackage{amsmath}
\usepackage{amssymb}
\usepackage{graphicx}%
\newtheorem{theorem}{Theorem}

\newtheorem{corollary}[theorem]{Corollary}

\newtheorem{definition}[theorem]{Definition}
\newtheorem{example}[theorem]{Example}

\newtheorem{lemma}[theorem]{Lemma}

\newtheorem{proposition}[theorem]{Proposition}
\newtheorem{remark}[theorem]{Remark}

\newenvironment{proof}[1][Proof]{\noindent\textbf{#1.} }{\ \rule{0.5em}{0.5em}}
\begin{document}

\title{Observables as functions:\\Antonymous functions}
\author{Andreas D\"{o}ring\\IAMPh, Fachbereich Mathematik,\\J. W. Goethe-Universit\"{a}t Frankfurt, Germany\\{\normalsize {adoering@math.uni-frankfurt.de}}}
\date{October 13, 2005}
\maketitle

\begin{abstract}
Antonymous functions are real-valued functions on the Stone spectrum of a von
Neumann algebra $\mathcal{R}$. They correspond to the self-adjoint operators
in $\mathcal{R}$, which are interpreted as observables in quantum physics.
Antonymous functions turn out to be generalized Gelfand transforms, related to
de Groote's observable functions.

\end{abstract}

\section{Introduction}

The \textbf{Stone spectrum} $\mathcal{Q(R)}$ of a unital von Neumann algebra
$\mathcal{R}\subseteq\mathcal{L(H)}$ is defined as the space of maximal filter
bases\footnote{A subset $\mathcal{F}$ of elements of a lattice $\mathbb{L}$
with zero element $0$ is a \textbf{filter base} if (i) $0\neq\mathcal{F}$ and
(ii) for all $a,b\in\mathcal{F}$, there is a $c\in\mathcal{F}$ such that
$c\leq a\wedge b$.} (or, equivalently, maximal dual ideals\footnote{A subset
$\mathcal{D}$ of elements of a lattice $\mathbb{L}$ with\ zero element $0$ is
a \textbf{dual ideal} if (i) $0\notin\mathcal{D}$, (ii) $a,b\in\mathcal{D}$
implies $a\wedge b\in\mathcal{D}$ and (iii) $a\in\mathcal{D}$ and $b\geq a$
imply $b\in\mathcal{D}$.} \cite{Bir73}) in the projection lattice
$\mathcal{P(R)}$ of $\mathcal{R}$ \cite{deG05}. The sets%
\[
\mathcal{Q}_{P}(\mathcal{R}):=\{\mathfrak{B}\in\mathcal{Q(R)}\ |\ P\in
\mathfrak{B}\},\quad P\in\mathcal{P(R)}%
\]
form the base of a topology on the Stone spectrum $\mathcal{Q(R)}$ such that
$\mathcal{Q(R)}$ becomes a zero-dimensional, completely regular Hausdorff
space. The sets $\mathcal{Q}_{P}\mathcal{(R)}$ are closed-open. If the von
Neumann algebra $\mathcal{R}$ is abelian, then the Stone spectrum
$\mathcal{Q(R)}$ is homeomorphic to the Gelfand spectrum $\Omega(\mathcal{R})$
of $\mathcal{R}$. For an arbitrary non-abelian unital von Neumann algebra
$\mathcal{R}$, the Stone spectrum can hence be regarded as a
\emph{non-commutative generalization of the Gelfand spectrum}. The elements
$\mathfrak{B}$ of the Stone spectrum $\mathcal{Q(R)}$ are called
\textbf{quasipoints}.

\ 

In this article, we show that for each self-adjoint operator $A\in
\mathcal{R}_{sa}$, there is a real-valued function $g_{A}:\mathcal{Q(R)}%
\rightarrow\mathbb{R}$ on the Stone spectrum $\mathcal{Q(R)}$ with the
following properties: (i) the image of $g_{A}$ is the spectrum of $A$, (ii)
$g_{A}$ is continuous and (iii) if $\mathcal{R}$ is abelian, then $g_{A}$
coincides with the Gelfand transform $\widehat{A}$ of $A$. The function
$g_{A}$, called the \textbf{function antonymous with }$A$ (or the antonymous
function of $A$), can hence be regarded as a \emph{generalized Gelfand
transform} of the self-adjoint operator $A\in\mathcal{R}_{sa}$.

\ 

Antonymous functions are related to \emph{observable functions} introduced by
de Groote and have analogous properties, compare \cite{deG05b}. Together,
these functions give a novel view on quantum observables. Each physical
observable, represented by a self-adjoint operator $A$, corresponds to a pair
of continuous functions on the Stone spectrum $\mathcal{Q(R)}$ of the von
Neumann algebra $\mathcal{R}$.

\ 

In section \ref{_SMotivation}, we will arrive at the definition of antonymous
functions by a motivation from physics. This will give a clear physical
interpretation of both antonymous and observable functions, which was lacking
up to now. Antonymous and observable functions show up as the lower resp.
upper integration limit in the integral defining physical expectation values.
The relation between antonymous and observable functions is shown.
Additionally, we give a presheaf construction leading to the definition of
antonymous functions.

\ 

The properties (i)--(iii) mentioned above are demonstrated in section
\ref{_SSomePropsOfAntonFcts}, using some results from \cite{deG05b}. Some
differences between antonymous functions and observable functions are shown.

\subsection{Conventions}

Throughout, $\mathcal{R}$ denotes a unital von Neumann algebra, given as a
subalgebra of the algebra $\mathcal{L(H)}$ of bounded linear operators on some
suitable Hilbert space $\mathcal{H}$. The real linear space of self-adjoint
operators in $\mathcal{R}$ is denoted by $\mathcal{R}_{sa}$, $\mathcal{P(R)}$
is the projection lattice of $\mathcal{R}$. A spectral family in
$\mathcal{P(R)}$ is a family $E=(E_{\lambda})_{\lambda\in\mathbb{R}}$ of
projections in $\mathcal{P(R)}$ such that (i) for all $\mu<\lambda$, $E_{\mu
}\leq E_{\lambda}$, (ii) $\bigwedge_{\lambda\in\mathbb{R}}E_{\lambda}=0$ and
(iii) $\bigvee_{\lambda\in\mathbb{R}}E_{\lambda}=I$. A spectral family $E$ is
bounded if there are $a,b\in\mathbb{R}$ such that $E_{\lambda}=0$ for all
$\lambda<a$ and $E_{\lambda}=I$ for all $\lambda>b$.

\ 

Let $A\in\mathcal{R}_{sa}$ be self-adjoint. The spectral theorem shows that
there is a bounded spectral family $E^{A}=(E_{\lambda}^{A})_{\lambda
\in\mathbb{R}}$ in $\mathcal{P(R)}$ such that $A=\int_{-|A|}^{|A|}\lambda
dE_{\lambda}$ (see e.g. \cite{KadRin97}). In order to make the spectral family
unique, one often additionally requires $E^{A}$ to be \emph{right-continuous},
i.e. for all $\lambda\in\mathbb{R}$,%
\[
\underset{\varepsilon\rightarrow+0}{\text{s-lim}}\ E_{\lambda+\varepsilon}%
^{A}=E_{\lambda}^{A},\quad\bigwedge_{\mu>\lambda}E_{\mu}^{A}=E_{\lambda}^{A}.
\]

We will denote right-continuous spectral families by $E$ (or $E^{A}$)
throughout. Alternatively, one can also require \emph{left-continuity}. Let
$F^{A}=(F_{\lambda}^{A})_{\lambda\in\mathbb{R}}$ denote the (unique)
left-continuous spectral family of $A$, in particular, for all $\lambda
\in\mathbb{R}$,%
\[
\underset{\varepsilon\rightarrow+0}{\text{s-lim}}\ F_{\lambda-\varepsilon}%
^{A}=F_{\lambda}^{A},\quad\bigvee_{\mu<\lambda}F_{\mu}^{A}=F_{\lambda}^{A}.
\]

We will denote left-continuous spectral families by $F$ (or $F^{A}$)
throughout. Let $\mathcal{E(R)}$ denote the set of bounded right-continuous
spectral families in $\mathcal{P(R)}$, and let $\mathcal{F(R)}$ denote the set
of bounded left-continuous spectral families in $\mathcal{P(R)}$. We have a
bijection $\varphi:\mathcal{E(R)}\rightarrow\mathcal{F(R)}$, which is
obviously defined as follows: let $E\in\mathcal{E(R)}$ be a right-continuous
spectral family. Define a left-continuous spectral family $\varphi
(E)\in\mathcal{F(R)}$ by%
\[
\forall\lambda\in\mathbb{R}:\varphi(E)_{\lambda}:=\bigvee_{\mu<\lambda}E_{\mu
}.
\]
Conversely, let $F\in\mathcal{F(R)}$ be a left-continuous spectral family.
Define a right-continuous spectral family $\varphi^{-1}(F)\in\mathcal{E(R)}$
by%
\[
\forall\lambda\in\mathbb{R}:\varphi^{-1}(F)_{\lambda}:=\bigwedge_{\mu>\lambda
}F_{\mu}.
\]
Unless otherwise mentioned, spectral family will always mean left-continuous
spectral family.

\section{Motivation\label{_SMotivation}}

\subsection{Antonymous functions from physical expectation
values\label{_SubSAntonFctsFromPhysics}}

In physics, von Neumann algebras show up as the \emph{algebras of observables}
of quantum systems; the self-adjoint operators $A\in\mathcal{R}_{sa}$ are
\emph{quantum observables}.

\ 

Let $\mathcal{R}=\mathcal{L(H)}$. The \emph{expectation value} of an
observable $A\in\mathcal{L(H)}_{sa}$ in the pure state $\rho_{\mathbb{C}%
x}=P_{\mathbb{C}x}$ ($|x|=1$) is given by%
\begin{equation}
\left\langle Ax,x\right\rangle =tr(\rho_{\mathbb{C}x}A)=\int_{-|A|}%
^{|A|}\lambda\ d\left\langle E_{\lambda}^{A}x,x\right\rangle , \tag{1}%
\end{equation}
where $E^{A}=(E_{\lambda}^{A})_{\lambda\in\mathbb{R}}$ is the
(right-continuous) spectral family of $A$. (This can be found in textbooks on
quantum mechanics, see e.g. \cite{Emch84}.) Let $\mathcal{Q(L(H))}$ be the
Stone spectrum of $\mathcal{L(H)}$, that is, the space of maximal filter bases
in the projection lattice $\mathcal{P(L(H))}$. Let $\mathbb{C}x$ be the line
in Hilbert space defined by $x$, and let $P_{\mathbb{C}x}$ be the orthogonal
projection onto this line. There is a quasipoint $\mathfrak{B}_{\mathbb{C}x}$,
i.e. an element of the Stone spectrum $\mathcal{Q(L(H))}$, given by
\[
\mathfrak{B}_{\mathbb{C}x}:=\{P\in\mathcal{P(L(H))}\ |\ P\geq P_{\mathbb{C}%
x}\}.
\]
This obviously is a maximal filter base in $\mathcal{P(L(H))}$. Such a
quasipoint $\mathfrak{B}_{\mathbb{C}x}$ is an isolated point of
$\mathcal{Q(L(H))}$ and is called an \textbf{atomic quasipoint}. Let
\begin{align*}
f_{A}:\mathcal{Q(L(H))}  &  \longrightarrow\mathbb{R}\\
\mathfrak{B}  &  \longmapsto\inf\{\lambda\in\mathbb{R}\ |\ E_{\lambda}^{A}%
\in\mathfrak{B}\}
\end{align*}
be the observable function of $A$, where $E^{A}=(E_{\lambda}^{A})_{\lambda
\in\mathbb{R}}$ is the right-continuous spectral family of $A$ \cite{deG05b}.
Since we have%
\begin{align*}
f_{A}(\mathfrak{B}_{\mathbb{C}x})  &  =\inf\{\lambda\in\mathbb{R}%
\ |\ E_{\lambda}^{A}\in\mathfrak{B}_{\mathbb{C}x}\}\\
&  =\inf\{\lambda\in\mathbb{R}\ |\ P_{\mathbb{C}x}\leq E_{\lambda}^{A}\},
\end{align*}
$P_{\mathbb{C}x}\leq E_{\mu}^{A}$ holds for all $\mu>f_{A}(\mathfrak{B}%
_{\mathbb{C}x})$, so these values do not contribute to the integral and we can
write%
\begin{equation}
\left\langle Ax,x\right\rangle =tr(\rho_{\mathbb{C}x}A)=\int_{-|A|}%
^{f_{A}(\mathfrak{B}_{\mathbb{C}x})}\lambda\ d\left\langle E_{\lambda}%
^{A}x,x\right\rangle . \tag{2}%
\end{equation}
The physical meaning of the observable function $f_{A}$, evaluated at
$\mathfrak{B}_{\mathbb{C}x}\in\mathcal{Q(L(H))}$, hence is the following:
$f_{A}(\mathfrak{B}_{\mathbb{C}x})$ is the \emph{largest possible measurement
result} one can obtain for the observable $A$ when the system is in the pure
state $\rho_{\mathbb{C}x}$.

\ 

If the physical system is not in a pure state $\rho_{\mathbb{C}x}$, but in a
mixed state $\rho$, we use the fact that such a mixed state is the convex
combination of pure states, $\rho=\sum_{j}a_{j}\rho_{\mathbb{C}x_{j}}$. (The
unit vectors $x_{0},x_{1},...$ are chosen orthogonal.) Since the trace is
linear, for all $A\in\mathcal{R}_{sa}$ we have%
\[
tr(\rho A)=\sum_{j}a_{j}tr(\rho_{\mathbb{C}x_{j}}A),
\]
so the calculation of every expectation value $tr(\rho A)$ can be reduced to
the simple situation (2).

\ 

This suggests to search for a similar description for the lower integration
limit in (2). It can usually be taken larger than $-|A|$: we do not have to
consider those values of $\lambda$ for which $E_{\lambda}^{A}x=0$ holds.
Geometrically, for those $\lambda$ the line $\mathbb{C}x$ is contained in the
orthogonal complement of $U_{E_{\lambda}^{A}}$, the closed subspace
$E_{\lambda}^{A}$ projects onto. We define a function $g_{A}:\mathbb{P}%
(\mathcal{H})\rightarrow\mathbb{R}$ on projective Hilbert space to encode
this:%
\begin{align*}
g_{A}(\mathbb{C}x)  &  :=\sup\{\lambda\in\mathbb{R}\ |\ \mathbb{C}x\perp
U_{E_{\lambda}^{A}}\}\\
&  =\sup\{\lambda\in\mathbb{R}\ |\ P_{\mathbb{C}x}E_{\lambda}^{A}=0\}\\
&  =\sup\{\lambda\in\mathbb{R}\ |\ P_{\mathbb{C}x}\leq I-E_{\lambda}^{A}\}\\
&  =\sup\{\lambda\in\mathbb{R}\ |\ I-E_{\lambda}^{A}\in\mathfrak{B}%
_{\mathbb{C}x}\}.
\end{align*}
Let $F^{A}$ denote the left-continuous spectral family of $A$. Since
$\sup\{\lambda\in\mathbb{R}\ |\ I-E_{\lambda}^{A}\in\mathfrak{B}_{\mathbb{C}%
x}\}=\sup\{\lambda\in\mathbb{R}\ |\ I-F_{\lambda}^{A}\in\mathfrak{B}%
_{\mathbb{C}x}\}$, we also have%
\[
g_{A}(\mathbb{C}x)=\sup\{\lambda\in\mathbb{R}\ |\ I-F_{\lambda}^{A}%
\in\mathfrak{B}_{\mathbb{C}x}\}.
\]
(While this is not important currently, left-continuous spectral families turn
out to be more convenient in some situations.) The definition of $g_{A}$
allows a natural generalization:

\begin{definition}
\label{DAntonymusFunction}Let $\mathcal{R}$ be a von Neumann algebra, and let
$A\in\mathcal{R}_{sa}$ be a self-adjoint operator with spectral family $F^{A}%
$. The \textbf{function antonymous with }$A$ (or the \textbf{antonymous
function of }$A$) is the function%
\begin{align*}
g_{A}:\mathcal{Q(R)}  &  \longrightarrow\mathbb{R}\\
\mathfrak{B}  &  \longmapsto\sup\{\lambda\in\mathbb{R}\ |\ I-F_{\lambda}%
^{A}\in\mathfrak{B}\}.
\end{align*}
The set of antonymous functions of $\mathcal{R}$ is denoted by $\mathcal{A(R)}%
$.
\end{definition}

The name \textquotedblleft antonymous function\textquotedblright\ stems from
the fact that we consider the supremum of those real values $\lambda$ for
which $I-F_{\lambda}^{A}$ is contained in $\mathfrak{B}$, and $I-F_{\lambda
}^{A}$ is \textquotedblleft opposite to\textquotedblright\ or
\textquotedblleft antonymous with\textquotedblright\ $F^{A}$ (and hence with
$A$).

\ 

Just as wanted, the value $g_{A}(\mathfrak{B}_{\mathbb{C}x})$ obviously is the
\emph{smallest possible measurement result} one can obtain for the observable
$A$ when the system is in the pure state $\rho_{\mathbb{C}x}$ (where, as
before, $|x|=1$). We obtain for the expectation value of the observable
$A\in\mathcal{R}_{sa}$ in the pure state $\rho_{\mathbb{C}x}$:%
\begin{equation}
\left\langle Ax,x\right\rangle =tr(\rho_{\mathbb{C}x}A)=\int_{g_{A}%
(\mathfrak{B}_{\mathbb{C}x})}^{f_{A}(\mathfrak{B}_{\mathbb{C}x})}%
\lambda\ d\left\langle E_{\lambda}^{A}x,x\right\rangle . \tag{3}%
\end{equation}

There is a close relationship between antonymous and observable functions: If
$A\in\mathcal{R}_{sa}$ is a self-adjoint operator with right-continuous
spectral family $E^{A}$, then $(I-E_{(1-\lambda)-}^{A})_{\lambda\in\mathbb{R}%
}$ is the right-continuous spectral family of the operator $I-A$. Here,
$I-E_{(1-\lambda)-}^{A}=\bigwedge_{\mu>\lambda}(I-E_{1-\mu}^{A})$. Now the
observable function $f_{I-A}$ of the operator $I-A$ is%
\begin{align*}
f_{I-A}(\mathfrak{B})  &  =\inf\{\lambda\in\mathbb{R}\ |\ I-E_{(1-\lambda
)-}^{A}\in\mathfrak{B}\}\\
&  =\inf\{1-\lambda\in\mathbb{R}\ |\ I-E_{\lambda}^{A}\in\mathfrak{B}\}\\
&  =1-\sup\{\lambda\in\mathbb{R}\ |\ I-E_{\lambda}^{A}\in\mathfrak{B}\}\\
&  =1-g_{A}(\mathfrak{B}).
\end{align*}
A simple result on observable functions (lemma 2.1 in \cite{deG05b}) shows
that $1-f_{I-A}=-f_{-A}$. Analogously, $1-g_{I-A}=-g_{-A}$, see Lemma
\ref{Lg_(A+aI)=a+g_A} below. It is interesting that the smallest possible
measurement result $g_{A}(\mathfrak{B}_{\mathbb{C}x})$ and the largest
possible measurement result $f_{A}(\mathfrak{B}_{\mathbb{C}x})$ for the
observable $A\in\mathcal{R}_{sa}$ in the state $\rho_{\mathbb{C}x}$ are thus
related by%
\begin{align*}
g_{A}(\mathfrak{B}_{\mathbb{C}x})  &  =1-f_{I-A}(\mathfrak{B}_{\mathbb{C}%
x})=-f_{-A}(\mathfrak{B}_{\mathbb{C}x}),\\
f_{A}(\mathfrak{B}_{\mathbb{C}x})  &  =1-g_{I-A}(\mathfrak{B}_{\mathbb{C}%
x})=-g_{-A}(\mathfrak{B}_{\mathbb{C}x}).
\end{align*}

While many properties of antonymous functions can be derived from those of
observable functions (and vice versa) using these relations, others are not
obvious. Moreover, there is no apparent reason to consider the function
$1-f_{I-A}$, while the expression (3) for the expectation values gives a clear
physical interpretation of antonymous and observable functions. The antonymous
function $g_{A}$ and the observable function $f_{A}$ represent two different
aspects of the observable $A$.

\subsection{Antonymous functions from a presheaf construction}

The definition of antonymous functions can also be motivated from a presheaf
construction, as will be shown here. As before, let $\mathcal{R}$ denote a
unital von Neumann algebra, and let $\mathcal{S(R)}$ denote the category of
subalgebras of $\mathcal{R}$ of the form $P\mathcal{R}P$, $P\in\mathcal{P(R)}%
$. A morphism $P\mathcal{R}P\rightarrow Q\mathcal{R}Q$ between objects
$P\mathcal{R}P$, $Q\mathcal{R}Q$ of $\mathcal{S(R)}$ exists if and only if
$P\leq Q$. Then the morphism $P\mathcal{R}P\rightarrow Q\mathcal{R}Q$ simply
is the inclusion.

\begin{definition}
Let $\mathcal{R}$ be a unital von Neumann algebra. An \textbf{opposite
spectral family in }$\mathcal{P(R)}$ is a family $G=(G_{\lambda})_{\lambda
\in\mathbb{R}}$ of projections in $\mathcal{P(R)}$, indexed by the real
numbers, such that

\begin{enumerate}
\item[(i)] $\forall\lambda,\mu\in\mathbb{R}$, $\lambda<\mu:G_{\lambda}\geq
G_{\mu},$

\item[(ii)] $\bigwedge_{\mu<\lambda}G_{\mu}=G_{\lambda},$

\item[(iii)] $\bigwedge_{\lambda\in\mathbb{R}}G_{\lambda}=0,\quad
\bigvee_{\lambda\in\mathbb{R}}G_{\lambda}=I$.
\end{enumerate}

An opposite spectral family $G$ is called \textbf{bounded} if there exist
$a,b\in\mathbb{R}$ such that

\begin{enumerate}
\item[(iv)] $\forall\lambda<a:G_{\lambda}=I,\quad\forall\lambda>b:G_{\lambda
}=0$.
\end{enumerate}

Let $\mathcal{F}^{o}\mathcal{(R)}$ denote the set of bounded opposite spectral
families in $\mathcal{P(R)}$.
\end{definition}

If $G$ is a (bounded) opposite spectral family, then $I-G=(I-G_{\lambda
})_{\lambda\in\mathbb{R}}$ is a (bounded) spectral family. In particular,
$I-G$ is left-continuous. Conversely, let $F^{A}\in\mathcal{F(R)}$ be a
bounded spectral family in $\mathcal{P(R)}$, corresponding to a self-adjoint
operator $A\in\mathcal{R}_{sa}$. Then $I-F^{A}=(I-F_{\lambda}^{A})_{\lambda
\in\mathbb{R}}$ is a bounded opposite spectral family. This (and the spectral
theorem) shows that there are bijections%
\[
\mathcal{F}^{o}\mathcal{(R)}\simeq\mathcal{F(R)}\simeq\mathcal{R}_{sa}%
\]
between the sets of bounded opposite spectral families in $\mathcal{P(R)}$,
bounded spectral families in $\mathcal{P(R)}$ and self-adjoint operators in
$\mathcal{R}$. Any bounded opposite spectral family $G\in\mathcal{F}%
^{o}\mathcal{(R)}$ hence is of the form $G=I-F^{A}$ for some self-adjoint
operator $A\in\mathcal{R}_{sa}$.

\ 

Let $P\leq Q$ be projections in $\mathcal{P(R)}$. We will now define a
restriction mapping%
\begin{align*}
\rho_{P}^{Q}:\mathcal{F}^{o}(Q\mathcal{R}Q)  &  \longrightarrow\mathcal{F}%
^{o}(P\mathcal{R}P)\\
I-F^{A}  &  \longmapsto(I-F^{A})^{P}%
\end{align*}
between bounded opposite spectral families: let $I-F^{A}$ be a bounded
opposite spectral family in $\mathcal{P}(Q\mathcal{R}Q)$ (of course, the unit
$I$ in $Q\mathcal{R}Q$ is $I=Q$), and let%
\[
(I-F^{A})^{P}=((I-F^{A})_{\lambda}^{P})_{\lambda\in\mathbb{R}}:=((I-F_{\lambda
}^{A})\wedge P)_{\lambda\in\mathbb{R}}.
\]

\begin{lemma}
$(I-F^{A})^{P}$ is a bounded opposite spectral family in $\mathcal{P}%
(P\mathcal{R}P)$.
\end{lemma}

\begin{proof}
$(I-F^{A})^{P}$ fulfills all the defining conditions of a bounded spectral family:

\begin{enumerate}
\item[(i)] for all $\lambda<\mu$, $(I-F_{\lambda}^{A})\wedge P\geq(I-F_{\mu
}^{A})\wedge P$,

\item[(ii)] $\bigwedge_{\mu<\lambda}((I-F_{\mu}^{A})\wedge P)=P\wedge
\bigwedge_{\mu<\lambda}(I-F_{\mu}^{A})=P\wedge(I-\bigvee_{\mu<\lambda}F_{\mu
}^{A})=P\wedge(I-F_{\lambda}^{A}),$

\item[(iv)] (implies (iii), also) $I-F^{A}$ is bounded, so there exist
$a,b\in\mathbb{R}$ such that $I-F_{\lambda}^{A}=I$ for all $\lambda<a$ and
$I-F_{\lambda}^{A}=0$ for all $\lambda>b$. Clearly, we have $(I-F_{\lambda
}^{A})\wedge P=P$ for all $\lambda<a$ and $(I-F_{\lambda}^{A})\wedge P=0$ for
all $\lambda>b$, so $(I-F^{A})^{P}$ is bounded, too.
\end{enumerate}
\end{proof}

\begin{corollary}
The sets $\mathcal{F}^{o}(Q\mathcal{R}Q)$, $Q\in\mathcal{P(R)}$ of bounded
opposite spectral families, together with the restriction mappings $\rho
_{P}^{Q}:$\ $\mathcal{F}^{o}(Q\mathcal{R}Q)\rightarrow\mathcal{F}%
^{o}(P\mathcal{R}P)$ for any $P,Q\in\mathcal{P(R)}$ such that $P<Q$ form a
presheaf on the category $\mathcal{S(R)}$ of subalgebras of $\mathcal{R}$. It
will be called the \textbf{opposite spectral presheaf} on the category
$\mathcal{S(R)}$.
\end{corollary}

Now let $\mathcal{R}=\mathcal{L(H)}$. The idea is to restrict a bounded
opposite spectral family $I-F^{A}\subseteq\mathcal{P(L(H))}$ to a
one-dimensional subspace: let $P=P_{\mathbb{C}x}$ be the projection onto the
one-dimensional subspace $\mathbb{C}x\subset\mathcal{H}$, and let
$(I-F^{A})\wedge P_{\mathbb{C}x}$ be the restriction of $I-F^{A}$ to
$P_{\mathbb{C}x}\mathcal{L(H)}P_{\mathbb{C}x}=\mathbb{C}P_{\mathbb{C}x}$. Then
$(I-F^{A})\wedge P_{\mathbb{C}x}$ is given by%
\[
((I-F^{A})\wedge P_{\mathbb{C}x})_{\lambda}=(I-F_{\lambda}^{A})\wedge
P_{\mathbb{C}x}=\left\{
\begin{tabular}
[c]{ll}%
$P_{\mathbb{C}x}$ & for $\lambda\leq d(\mathbb{C}x)$\\
$0$ & for $\lambda>d(\mathbb{C}x)$%
\end{tabular}
\right.
\]
for some real parameter $d(\mathbb{C}x)$. Let $\mathfrak{B}_{\mathbb{C}x}$ be
the atomic quasipoint $\mathfrak{B}_{\mathbb{C}x}=\{P\in\mathcal{P(L(H))}%
\ |\ P\geq P_{\mathbb{C}x}\}$. The parameter $d(\mathbb{C}x)$ is given by%
\begin{align*}
d(\mathbb{C}x)  &  =\sup\{\lambda\in\mathbb{R}\ |\ (I-F_{\lambda}^{A})\wedge
P_{\mathbb{C}x}=P_{\mathbb{C}x}\}\\
&  =\sup\{\lambda\in\mathbb{R}\ |\ P_{\mathbb{C}x}\leq I-F_{\lambda}^{A}\}\\
&  =\sup\{\lambda\in\mathbb{R}\ |\ I-F_{\lambda}^{A}\in\mathfrak{B}%
_{\mathbb{C}x}\}\\
&  =g_{A}(\mathfrak{B}_{\mathbb{C}x}),
\end{align*}
so the restriction of the bounded opposite spectral family $I-F^{A}$ to
one-dimensional subspaces leads to the definition of the antonymous function
$g_{A}$ of $A\in\mathcal{R}_{sa}$. This is similar to the situation for
observable functions, which were originally defined using the restriction of
spectral families to one-dimensional subspaces \cite{deG01}. The definition of
opposite spectral families was made to capture the intuitive idea that the
antonymous function $g_{A}$ represents information that is contained in the
\textquotedblleft opposite\textquotedblright\ $I-F^{A}$ of the spectral family
$F^{A}$.

\section{Some properties of antonymous functions\label{_SSomePropsOfAntonFcts}%
}

In this section, we clarify some of the properties of antonymous functions. In
particular, we will show that antonymous functions are generalized Gelfand
transforms in the sense that for an abelian von Neumann algebra, the
antonymous function $g_{A}$ and $\widehat{A}$, the Gelfand transform of a
self-adjoint operator $A$, coincide.

\subsection{Basic properties}

In this subsection, we draw on some results found for observable functions. To
make this paper reasonably self-contained, we give full proofs. Compare
\cite{deG05b}, in particular, for the proofs of Props. \ref{Pimg_A=spA},
\ref{Pg_AForFiniteRealLinComb}, Lemma \ref{Lg_(A+aI)=a+g_A} and Thm.
\ref{Tg_AContinuous}.

\ 

We start with the important example of the antonymous function of a projection
$P$, which turns out to be a characteristic function:

\begin{example}
\label{Eg_PForPProj}Let $P\in\mathcal{P(R)}\backslash\{0\}$ be a non-zero
projection in the von Neumann algebra $\mathcal{R}$. The spectral family of
$P$ is given by%
\[
F_{\lambda}^{P}=\left\{
\begin{tabular}
[c]{ll}%
$0$ & for $\lambda\leq0$\\
$I-P$ & for $0<\lambda\leq1$\\
$I$ & for $\lambda>1$.
\end{tabular}
\ \right.
\]
If $\mathfrak{B}\in\mathcal{Q}_{P}(\mathcal{R})$, i.e. $P\in\mathfrak{B}$,
then%
\[
g_{P}(\mathfrak{B})=\sup\{\lambda\in\mathbb{R}\ |\ I-F_{\lambda}^{P}%
\in\mathfrak{B}\}=1,
\]
while for a quasipoint $\mathfrak{B}\notin\mathcal{Q}_{P}(\mathcal{R})$ we
have $g_{P}(\mathfrak{B})=0$, so%
\[
g_{P}=\chi_{\mathcal{Q}_{P}(\mathcal{R})}.
\]
The antonymous function $g_{P}$ of a projection $P$ is continuous, since the
sets $\mathcal{Q}_{P}\mathcal{(R)}$ are closed-open.
\end{example}

(The observable function $f_{P}$ of a projection $P$ is $f_{P}=1-\chi
_{\mathcal{Q}_{I-P}(\mathcal{R})}$.) The image of the antonymous function
$g_{P}$ is $\{0,1\}$, and this coincides with the spectrum $\operatorname*{sp}%
P$ of $P$. (If $P=I$, then $\operatorname*{im}g_{I}=\operatorname*{im}%
\chi_{\mathcal{Q}_{I}(\mathcal{R})}=\operatorname*{im}\chi_{\mathcal{Q(R)}%
}=\{1\}=\operatorname*{sp}I$.) We now show that $\operatorname*{im}%
g_{A}=\operatorname*{sp}A$ holds not just for projections, but for all
self-adjoint operators $A$:

\begin{proposition}
\label{Pimg_A=spA}Let $\mathcal{R}$ be a von Neumann algebra, and let
$A\in\mathcal{R}_{sa}$. Then $\operatorname*{im}g_{A}=\operatorname*{sp}A$.
\end{proposition}

\begin{proof}
$\operatorname*{sp}A$ consists of those $\lambda\in\mathbb{R}$ such that
$F^{A}$ is non-constant on every open neighbourhood of $\lambda$. Let
$\lambda_{0}\in\operatorname*{im}g_{A}$, but $\lambda_{0}\notin
\operatorname*{sp}A$. Then there is some $\varepsilon>0$ such that%
\[
\forall\lambda\in\:]\lambda_{0}-\varepsilon,\lambda_{0}+\varepsilon
\lbrack\;:F_{\lambda}^{A}=F_{\lambda_{0}}^{A}.
\]
Let $\mathfrak{B}\in\overset{-1}{g_{A}}(\lambda_{0})$. From the definition of
$g_{A}$, we obtain $g_{A}(\mathfrak{B})\geq\lambda_{0}+\varepsilon$, which is
a contradiction. This implies%
\[
\operatorname*{im}g_{A}\subseteq\operatorname*{sp}A.
\]
Conversely, let $\lambda_{0}\in\operatorname*{sp}A$. We have to find a
quasipoint $\mathfrak{B}\in\mathcal{Q(R)}$ such that $g_{A}(\mathfrak{B}%
)=\lambda_{0}$.

\emph{Case (i):} $F_{\lambda}^{A}<F_{\lambda_{0}}^{A}$ for all $\lambda
<\lambda_{0}$. We choose some $\mathfrak{B}$ that contains all projections
$F_{\lambda_{0}}^{A}-F_{\lambda}^{A}$ for $\lambda<\lambda_{0}$. Then all the
projections $I-F_{\lambda}^{A}$ for $\lambda<\lambda_{0}$ are contained in
$\mathfrak{B}$, since $I-F_{\lambda}^{A}\geq F_{\lambda_{0}}^{A}-F_{\lambda
}^{A}$ and $\mathfrak{B}$ is a dual ideal in $\mathcal{P(R)}$. Moreover,
$F_{\lambda_{0}}^{A}\in\mathfrak{B}$, since $F_{\lambda_{0}}^{A}%
>F_{\lambda_{0}}^{A}-F_{\lambda}^{A}$ for $\lambda<\lambda_{0}$, so
$I-F_{\lambda_{0}}^{A}\notin\mathfrak{B}$ (the projections $F_{\lambda_{0}%
}^{A}$ and $I-F_{\lambda_{0}}^{A}$ cannot both be contained in a quasipoint
$\mathfrak{B}$, since $\mathfrak{B}$ is a filter base and hence does not
contain the zero projection $0=F_{\lambda_{0}}^{A}\wedge(I-F_{\lambda_{0}}%
^{A})$). This implies $g_{A}(\mathfrak{B})=\lambda_{0}$.

\emph{Case (ii)} (the two cases are not mutually exclusive): there is a
decreasing sequence $(\lambda_{n})_{n\in\mathbb{N}}$ such that $\lambda
_{0}=\lim_{n\rightarrow\infty}\lambda_{n}$ and $F_{\lambda_{n+1}}%
^{A}<F_{\lambda_{n}}^{A}$ for all $n$. Take a quasipoint $\mathfrak{B}$ that
contains all projections $F_{\mu}^{A}-F_{\lambda_{0}}^{A}$ for $\mu
>\lambda_{0}$. Then $\mathfrak{B}$ contains $I-F_{\lambda_{0}}^{A}$ and all
projections $F_{\mu}^{A}$ for $\mu>\lambda_{0}$, so $I-F_{\mu}^{A}%
\notin\mathfrak{B}$ for $\mu>\lambda_{0}$.
\end{proof}

\ 

Let $A\in\mathcal{R}_{sa}$ be self-adjoint, $f_{A}$ its observable function
and $g_{A}$ its antonymous function. We saw in section \ref{_SMotivation} that
for the von Neumann algebra $\mathcal{R}=\mathcal{L(H)}$, $g_{A}%
(\mathfrak{B}_{\mathbb{C}x})\leq f_{A}(\mathfrak{B}_{\mathbb{C}x})$ holds for
all atomic quasipoints $\mathfrak{B}_{\mathbb{C}x}\in\mathcal{Q(L(H))}$. Not
surprisingly, this also holds more generally:

\begin{proposition}
Let $\mathcal{R}$ be a von Neumann algebra, and let $A\in\mathcal{R}_{sa}$
with antonymous function $g_{A}$ and observable function $f_{A}$. For all
quasipoints $\mathfrak{B}\in\mathcal{Q(R)}$, we have $g_{A}(\mathfrak{B})\leq
f_{A}(\mathfrak{B})$.
\end{proposition}

\begin{proof}
Let $E^{A}$ be the right-continuous spectral family of $A$, and let $F^{A}$ be
the left-continuous one. Then we have%
\begin{align*}
g_{A}(\mathfrak{B})  &  =\sup\{\lambda\in\mathbb{R}\ |\ I-F_{\lambda}^{A}%
\in\mathfrak{B}\}\\
&  =\sup\{\lambda\in\mathbb{R}\ |\ I-E_{\lambda}^{A}\in\mathfrak{B}\}.
\end{align*}
Assume that there is some quasipoint $\mathfrak{B}\in\mathcal{Q(R)}$ such that
$g_{A}(\mathfrak{B})>f_{A}(\mathfrak{B})$. Choose some $\lambda_{0}\in
\ ]f_{A}(\mathfrak{B}),g_{A}(\mathfrak{B})[$. Then we have $g_{A}%
(\mathfrak{B})=\sup\{\lambda\in\mathbb{R}\ |\ I-E_{\lambda}^{A}\in
\mathfrak{B}\}>\lambda_{0}$, so the quasipoint $\mathfrak{B}$ must contain a
projection $I-E_{\lambda_{1}}^{A}$ for $\lambda_{1}>\lambda_{0}$. $E^{A}$ is a
spectral family, so $I-E_{\lambda_{1}}^{A}\leq I-E_{\lambda_{0}}^{A}$. Since
$\mathfrak{B}$ is a dual ideal in $\mathcal{P(R)}$, we have $I-E_{\lambda_{0}%
}^{A}\in\mathfrak{B}$.

On the other hand, $f_{A}(\mathfrak{B})=\inf\{\lambda\in\mathbb{R}%
\ |\ E_{\lambda}^{A}\in\mathfrak{B}\}<\lambda_{0}$, so $\mathfrak{B}$ contains
a projection $E_{\lambda_{2}}^{A}$ for $\lambda_{2}<\lambda_{0}$.
$E_{\lambda_{2}}^{A}\leq E_{\lambda_{0}}^{A}$ holds, so $E_{\lambda_{0}}%
^{A}\in\mathfrak{B}$. But $I-E_{\lambda_{0}}^{A},E_{\lambda_{0}}^{A}$ cannot
both be contained in a quasipoint $\mathfrak{B}$, since $\mathfrak{B}$ is a
filter base.
\end{proof}

\begin{example}
Let $\mathcal{R}$ be a non-abelian von Neumann algebra. If $P\in
\mathcal{P(R)}$ is a projection, then the strict inequality $g_{P}%
(\mathfrak{B})<f_{P}(\mathfrak{B})$ holds for all quasipoints $\mathfrak{B}$
neither contained in $\mathcal{Q}_{P}(\mathcal{R})$ nor in $\mathcal{Q}%
_{I-P}(\mathcal{R})$. Namely, $g_{P}(\mathfrak{B})=0<1=f_{P}(\mathfrak{B})$
for such a quasipoint.
\end{example}

While for non-abelian algebras $\mathcal{R}$, there are quasipoints neither
containing a given projection $P$ nor $I-P$, the next lemma, which is a
variant of a lemma from \cite{deG01}, shows that this cannot happen for
abelian von Neumann algebras:

\begin{lemma}
\label{LAbVNAQPContainsPOrI-P}Let $\mathcal{R}$ be an abelian von Neumann
algebra, and let $\mathfrak{B}\in\mathcal{Q(R)}$ be a quasipoint of
$\mathcal{R}$. For all projections $P\in\mathcal{P(R)}$, either $P\in
\mathfrak{B}$ or $I-P\in\mathfrak{B}$.
\end{lemma}

\begin{proof}
Clearly, at most one of the projections $P,I-P$ can be contained in a
quasipoint $\mathfrak{B}$, since $\mathfrak{B}$ is a filter base. In order to
show that at least one of the projections $P,I-P$ is contained in
$\mathfrak{B}$, we use the fact that the projection lattice $\mathcal{P(R)}$
of $\mathcal{R}$ is distributive, since $\mathcal{R}$ is abelian.
$P,(I-P)\notin\mathfrak{B}$ would imply
\[
\exists Q,R\in\mathfrak{B}:P\wedge Q=(I-P)\wedge R=0,
\]
so%
\begin{align*}
Q\wedge R  &  =(P\vee(I-P))\wedge Q\wedge R\\
&  =(P\wedge Q\wedge R)\vee((I-P)\wedge Q\wedge R)=0,
\end{align*}
contradicting $Q\wedge R\in\mathfrak{B}$.
\end{proof}

\ 

In other words, for an abelian von Neumann algebra $\mathcal{R}$, a quasipoint
$\mathfrak{B}$ is an \emph{ultrafilter} in the projection lattice
$\mathcal{P(R)}$.

\ \ 

The next proposition shows that for an abelian algebra $\mathcal{R}$, the
antonymous function $g_{A}$ and the observable function $f_{A}$ of a
self-adjoint operator $A\in\mathcal{R}$ coincide:

\begin{proposition}
\label{PAbVNAg_A=f_A}Let $\mathcal{R}$ be an abelian von Neumann algebra, and
let $A\in\mathcal{R}_{sa}$ with antonymous function $g_{A}$ and observable
function $f_{A}$. Then $g_{A}=f_{A}$.
\end{proposition}

\begin{proof}
Let $F^{A}$ be the left-continuous spectral family of $A$ and $E^{A}$ the
right-continuous one. Let $\mathfrak{B}\in\mathcal{Q(R)}$ be a quasipoint. We
have%
\begin{align*}
g_{A}(\mathfrak{B})  &  =\sup\{\lambda\in\mathbb{R}\ |\ I-F_{\lambda}^{A}%
\in\mathfrak{B}\}\\
&  =\sup\{\lambda\in\mathbb{R}\ |\ I-E_{\lambda}^{A}\in\mathfrak{B}\}.
\end{align*}
According to Lemma \ref{LAbVNAQPContainsPOrI-P}, for all $P\in\mathcal{P(R)}$,
either $P\in\mathfrak{B}$ or $I-P\in\mathfrak{B}$. Since $I-E_{\lambda}%
^{A}\notin\mathfrak{B}$ for all $\lambda>g_{A}(\mathfrak{B})$, we have
$E_{\lambda}^{A}\in\mathfrak{B}$ for all $\lambda>g_{A}(\mathfrak{B})$ (and
$E_{\mu}^{A}\notin\mathfrak{B}$ for all $\mu<g_{A}(\mathfrak{B})$) and thus%
\[
f_{A}(\mathfrak{B})=\inf\{\lambda\in\mathbb{R}\ |\ E_{\lambda}^{A}%
\in\mathfrak{B}\}=g_{A}(\mathfrak{B}).
\]

\end{proof}

\begin{lemma}
\label{Lg_(A+aI)=a+g_A}Let $A\in\mathcal{R}_{sa}$, and let $t\in\mathbb{R}$.
Then $g_{A+tI}=t+g_{A}$.
\end{lemma}

\begin{proof}
For all $\mathfrak{B}\in\mathcal{Q(R)}$, we have%
\begin{align*}
g_{A+tI}(\mathfrak{B})  &  =\sup\{\lambda\in\mathbb{R}\ |\ I-F_{\lambda
}^{A+tI}\in\mathfrak{B}\}\\
&  =\sup\{\lambda\in\mathbb{R}\ |\ I-F_{\lambda-t}^{A}\in\mathfrak{B}\}\\
&  =\sup\{t+\lambda\in\mathbb{R}\ |\ I-F_{\lambda}^{A}\in\mathfrak{B}\}\\
&  =t+\sup\{\lambda\in\mathbb{R}\ |\ I-F_{\lambda}^{A}\in\mathfrak{B}\}\\
&  =t+g_{A}(\mathfrak{B}).
\end{align*}

\end{proof}

\ 

We will now show that the antonymous function $g_{A}$ of a finite real linear
combination $A=\sum_{j=1}^{n}a_{j}P_{j}$ of pairwise orthogonal non-zero
projections $P_{1},\ldots,P_{n}\in\mathcal{P(R)}$ with non-zero real
coefficients $a_{1}\leq\ldots\leq a_{n}$ is a step function and hence
continuous. Let $P:=\sum_{j=1}^{n}P_{j}$. $0$ is contained in the spectrum of
the operator $A$ if and only if $P\neq I$. The following discussion is only
needed in the case $P\neq I$, but things also work trivially if $P=I$.

\ 

Let $k_{0}$ be the greatest index such that $a_{k_{0}}<0$. If there is no
$a_{k}<0$, then $k_{0}:=0$. For $k=1,...,n+1$, define%
\[
b_{k}:=\left\{
\begin{tabular}
[c]{ll}%
$a_{k}$ & for $k\leq k_{0}$\\
$0$ & for $k_{0}+1$\\
$a_{k-1}$ & for $k>k_{0}+1$,
\end{tabular}
\ \right.
\]%
\[
Q_{k}:=\left\{
\begin{tabular}
[c]{ll}%
$P_{k}$ & for $k\leq k_{0}$\\
$I-P$ & for $k=k_{0}+1$\\
$P_{k-1}$ & for $k>k_{0}+1$.
\end{tabular}
\ \right.
\]
Then $\sum_{k=1}^{n+1}Q_{k}=I$ and $A=\sum_{k=1}^{n+1}b_{k}Q_{k}$, where
$b_{1}\leq...\leq b_{n+1}$ and $b_{k_{0}+1}=0$, but $Q_{k_{0}+1}\neq0$ unless
$P=I$. Thus the spectral family $F^{A}$ of $A$ is:%
\[
F_{\lambda}^{A}=\left\{
\begin{tabular}
[c]{ll}%
$0$ & for $\lambda\leq b_{1}$\\
$Q_{1}+...+Q_{k}$ & for $b_{k}<\lambda\leq b_{k+1}$ ($k=1,...,n$)\\
$I$ & for $\lambda>b_{n+1}$.
\end{tabular}
\ \right.
\]

Let $Q_{0}:=0$. Each quasipoint $\mathfrak{B}\in\mathcal{Q(R)}$ is contained
in exactly one of the pairwise disjoint sets%
\begin{align*}
M_{1}  &  :=\mathcal{Q}_{I-Q_{0}}(\mathcal{R})\backslash\mathcal{Q}%
_{I-(Q_{0}+Q_{1})}(\mathcal{R})=\mathcal{Q(R)}\backslash\mathcal{Q}%
_{I-(Q_{0}+Q_{1})}(\mathcal{R}),\\
M_{2}  &  :=\mathcal{Q}_{I-(Q_{0}+Q_{1})}(\mathcal{R})\backslash
\mathcal{Q}_{I-(Q_{0}+Q_{1}+Q_{2})}(\mathcal{R}),\\
M_{3}  &  :=\mathcal{Q}_{I-(Q_{0}+Q_{1}+Q_{2})}(\mathcal{R})\backslash
\mathcal{Q}_{I-(Q_{0}+...+Q_{3})}(\mathcal{R}),\\
&  .\\
&  .\\
&  .\\
M_{n+1}  &  :=\mathcal{Q}_{I-(Q_{0}+...+Q_{n})}(\mathcal{R})\backslash
\mathcal{Q}_{I-(Q_{0}+...+Q_{n+1})}(\mathcal{R})\\
&  =\mathcal{Q}_{I-(Q_{0}+Q_{1}+...+Q_{n})}(\mathcal{R}).
\end{align*}
\ \bigskip The last equality holds since $Q_{0}+Q_{1}+...+Q_{n+1}=I$, so
$\mathcal{Q}_{I-(Q_{0}+Q_{1}+...+Q_{n+1})}(\mathcal{R})=\mathcal{Q}%
_{0}(\mathcal{R})=\varnothing$. Now, for $k=1,...,n+1$, we have%
\begin{align*}
\mathfrak{B}\in M_{k}\Longrightarrow g_{A}(\mathfrak{B})  &  =\sup\{\lambda
\in\mathbb{R}\ |\ I-F_{\lambda}^{A}\in\mathfrak{B}\}\\
&  =\sup\{\lambda\in\mathbb{R}\ |\ I-F_{\lambda}^{A}=I-(Q_{0}+Q_{1}%
+...+Q_{k-1})\}\\
&  =\sup\{\lambda\in\mathbb{R}\ |\ F_{\lambda}^{A}=Q_{0}+Q_{1}+...+Q_{k-1}\}\\
&  =b_{k}.
\end{align*}

Thus we obtain

\begin{proposition}
\label{Pg_AForFiniteRealLinComb}Let $P_{1},P_{2},\ldots,P_{n}\in
\mathcal{P(R)}$ be pairwise orthogonal non-zero projections and $A:=\sum
_{j=1}^{n}a_{j}P_{j}$ with real non-zero coefficients $a_{1}\leq...\leq a_{n}%
$. We write $A=\sum_{k=1}^{n+1}b_{k}Q_{k}$ with the $b_{k}$ and $Q_{k}$
defined as above. Let $Q_{0}:=0$. Then the antonymous function of $A$ is given
by%
\begin{equation}
g_{A}=\sum_{k=1}^{n+1}b_{k}\chi_{M_{k}}=\sum_{k=1}^{n+1}b_{k}\chi
_{\mathcal{Q}_{I-(Q_{0}+...+Q_{k-1})}(\mathcal{R})\backslash\mathcal{Q}%
_{I-(Q_{0}+...+Q_{k})}(\mathcal{R})}. \tag{$\ast$}%
\end{equation}

In particular, the antonymous function $g_{A}$ is continuous.
\end{proposition}

The antonymous function $g_{A}$ of a finite real linear combination
$A:=\sum_{j=1}^{n}a_{j}P_{j}$ of pairwise orthogonal projections hence is a
step function. Of course, the coefficient $b_{k_{0}+1}$ is $0$, so the summand
for $k_{0}+1$ can be left out from the sum ($\ast$). However, we must use the
projections $Q_{k}$ and not simply the $P_{j}$ in ($\ast$), since they show up
in the spectral family $F^{A}$. (This is not necessary if $P=\sum_{j=1}%
^{n}P_{j}=I$).

\ 

If we take $A=P\in\mathcal{P(R)}$, we get back the simple characteristic
function of example \ref{Eg_PForPProj}: we have $A=0\cdot(I-P)+1\cdot
P=\sum_{k=1}^{2}b_{k}Q_{k}$, where $b_{1}=0$, $b_{2}=1$, $Q_{1}=I-P$,
$Q_{2}=P$ and $Q_{0}:=0$. Inserting into ($\ast$) gives%
\begin{align*}
g_{P}  &  =b_{1}\chi_{\mathcal{Q}_{I-Q_{0}}(\mathcal{R})\backslash
\mathcal{Q}_{I-(Q_{0}+Q_{1})}(\mathcal{R})}+b_{2}\chi_{\mathcal{Q}%
_{I-(Q_{0}+Q_{1})}(\mathcal{R})\backslash\mathcal{Q}_{I-(Q_{0}+Q_{1}+Q_{2}%
)}(\mathcal{R})}\\
&  =b_{2}\chi_{\mathcal{Q}_{I-Q_{1}}(\mathcal{R})\backslash\mathcal{Q}%
_{0}(\mathcal{R})}\\
&  =\chi_{\mathcal{Q}_{P}(\mathcal{R})}.
\end{align*}

\begin{theorem}
\label{TAbvNAg_A=f_A}Let $\mathcal{R}$ be an abelian von Neumann algebra, and
let $A=\sum_{j=1}^{n}a_{j}P_{j}$ for pairwise orthogonal non-zero projections
$P_{1},...,P_{n}$ and non-zero real coefficients $a_{1},...,a_{n}$. Let
$b_{k}$, $Q_{k}$ ($k=1,...,n+1$) be defined as in Prop.
\ref{Pg_AForFiniteRealLinComb}, so $A=\sum_{k=1}^{n+1}b_{k}Q_{k}$. Then
\[
g_{A}=\sum_{k=1}^{n+1}b_{k}\chi_{\mathcal{Q}_{Q_{k}(\mathcal{R})}}=f_{A},
\]
where $g_{A}$ is the antonymous function and $f_{A}$ is the observable
function of $A$.
\end{theorem}

\begin{proof}
The second equality is shown in \cite{deG05b} (In fact, there is a minor
mistake in \cite{deG05b}: in prop. 2.6 and cor. 2.1, the projections $P_{j}$
and not the $Q_{k}$ are used in the expression for $f_{A}$, which is wrong
unless $P=\sum_{k=1}^{n}P_{k}=I$.) The equality of $g_{A}$ and $f_{A}$ is
shown in Prop. \ref{PAbVNAg_A=f_A}.

It might also be instructive to prove the first equality: if $\mathcal{R}$ is
abelian, then $\mathcal{P(R)}$ is distributive. Let $A=\sum_{k=1}^{n+1}%
b_{k}Q_{k}$ with $b_{k},Q_{k}$ as in Prop. \ref{Pg_AForFiniteRealLinComb}. Let
$Q_{0}:=0$. According to Lemma \ref{LAbVNAQPContainsPOrI-P}, a quasipoint
$\mathfrak{B}\in\mathcal{Q(R)}$ that does not contain $I-(Q_{0}+...+Q_{k})$
contains $Q_{0}+...+Q_{k}$. Thus a quasipoint $\mathfrak{B}\in\mathcal{Q}%
_{I-(Q_{0}+...+Q_{k-1})}(\mathcal{R})\backslash\mathcal{Q}_{I-(Q_{0}%
+...+Q_{k})}(\mathcal{R})$ contains $I-(Q_{0}+...+Q_{k-1})$ and $Q_{0}%
+...+Q_{k}$ and hence%
\[
I-(Q_{0}+...+Q_{k-1})\wedge(Q_{0}+...+Q_{k})=Q_{k}\in\mathfrak{B}.
\]
Conversely, even in the non-distributive case each $\mathfrak{B}\in
\mathcal{Q}_{Q_{k}}(\mathcal{R})$ contains the projection $I-(Q_{0}%
+...+Q_{k-1})$, but not the projection $I-(Q_{0}+...+Q_{k})$. So for abelian
$\mathcal{R}$, we have%
\[
\chi_{\mathcal{Q}_{I-(Q_{0}+\ldots+Q_{k-1})}(\mathcal{R})\setminus
\mathcal{Q}_{I-(Q_{0}+\ldots+Q_{k})}(\mathcal{R})}=\chi_{\mathcal{Q}_{Q_{k}%
}(\mathcal{R})}.
\]

\end{proof}

\ 

In order to show that the antonymous function $g_{A}$ of an arbitrary
self-adjoint operator $A\in\mathcal{R}_{sa}$ is continuous, we approximate
$g_{A}$ uniformly by the step functions from Prop.
\ref{Pg_AForFiniteRealLinComb}.

\begin{theorem}
\label{Tg_AContinuous}Let $\mathcal{R}$ be an arbitrary unital von Neumann
algebra, and let $A\in\mathcal{R}_{sa}$ be self-adjoint. Then the antonymous
function $g_{A}:\mathcal{Q(R)}\rightarrow\mathbb{R}$ of $A$ is continuous.
\end{theorem}

\begin{proof}
Let $m:=\min\operatorname*{sp}A$, $M:=\max\operatorname*{sp}A$ and
$\varepsilon>0$. Choose $\lambda_{0}\in\ ]m-\varepsilon,m[$, $\lambda_{n}%
\in\ ]M,M+\varepsilon\lbrack$, and $\lambda_{1},...,\lambda_{n-1}\in\ ]a,b[$
such that $\lambda_{k-1}<\lambda_{k}$ and $\lambda_{k}-\lambda_{k-1}%
<\varepsilon$ for all $k=1,...,n$. Moreover, choose $a_{k}\in\ ]\lambda
_{k-1},\lambda_{k}[$ for $k=1,...,n$ and define%
\[
A_{\varepsilon}:=\sum_{k=1}^{n}a_{k}(F_{\lambda_{k}}^{A}-F_{\lambda_{k-1}}%
^{A})=:\sum_{k=1}^{n}a_{k}P_{k},
\]
where $P_{k}:=F_{\lambda_{k}}^{A}-F_{\lambda_{k-1}}^{A}$. Let $P_{0}:=0$. The
spectral family of $A_{\varepsilon}$ is given by%
\[
F_{\lambda}^{A_{\varepsilon}}=\left\{
\begin{tabular}
[c]{ll}%
$0=F_{\lambda_{0}}^{A}$ & for $\lambda\leq a_{1}=\min\operatorname*{sp}%
A_{\varepsilon}$\\
$F_{\lambda_{k}}^{A}$ & for $a_{k}<\lambda\leq a_{k+1}$ ($k=1,...,n-1$)\\
$I=F_{\lambda_{n}}^{A}$ & for $\lambda>a_{n}=\max\operatorname*{sp}%
A_{\varepsilon}$,
\end{tabular}
\ \ \right.
\]
and from Prop. \ref{Pg_AForFiniteRealLinComb}, the antonymous function of
$g_{A_{\varepsilon}}$ is\footnote{Since $F_{\lambda_{0}}^{A}=0$ and
$F_{\lambda_{n}}^{A}=I$, we have $\sum_{k=1}^{n}P_{k}=I$ by construction.
Hence we do not have to mind if $0$ is a spectral value of $A$ or not and can
skip the definition of the $b_{k}$ and $Q_{k}$ ($k=1,...,n+1$) normally used
in Prop. \ref{Pg_AForFiniteRealLinComb}.}%
\begin{align*}
g_{A_{\varepsilon}}  &  =\sum_{k=1}^{n}a_{k}\chi_{\mathcal{Q}_{I-(P_{0}%
+...+P_{k-1})}(\mathcal{R})\backslash\mathcal{Q}_{I-(P_{0}+...+P_{k}%
)}(\mathcal{R})}\\
&  =\sum_{k=1}^{n}a_{k}\chi_{\mathcal{Q}_{I-F_{\lambda_{k-1}}^{A}}%
(\mathcal{R})\backslash\mathcal{Q}_{I-F_{\lambda_{k}}^{A}}(\mathcal{R})}.
\end{align*}

Each quasipoint $\mathfrak{B}\in\mathcal{Q(R)}$ is contained in exactly one of
the pairwise disjoint sets%
\begin{align*}
N_{1}  &  :=\mathcal{Q}_{I-F_{\lambda_{0}}^{A}}(\mathcal{R})\backslash
\mathcal{Q}_{I-F_{\lambda_{1}}^{A}}(\mathcal{R})=\mathcal{Q(R)}\backslash
\mathcal{Q}_{I-F_{\lambda_{1}}^{A}}(\mathcal{R}),\\
N_{2}  &  :=\mathcal{Q}_{I-F_{\lambda_{1}}^{A}}(\mathcal{R})\backslash
\mathcal{Q}_{I-F_{\lambda_{2}}^{A}}(\mathcal{R}),\\
&  .\\
&  .\\
&  .\\
N_{n}  &  :=\mathcal{Q}_{I-F_{\lambda_{n-1}}^{A}}(\mathcal{R})\backslash
\mathcal{Q}_{I-F_{\lambda_{n}}^{A}}(\mathcal{R})=\mathcal{Q}_{I-F_{\lambda
_{n-1}}^{A}}(\mathcal{R})\backslash\mathcal{Q}_{0}(\mathcal{R})\\
&  =\mathcal{Q}_{I-F_{\lambda_{n-1}}^{A}}(\mathcal{R}),
\end{align*}
so, for $k=1,...,n$, we obtain%
\begin{align*}
\mathfrak{B}\in N_{k}\Longrightarrow g_{A\varepsilon}(\mathfrak{B})  &
=\sup\{\lambda\in\mathbb{R}\ |\ I-F_{\lambda}^{A_{\varepsilon}}\in
\mathfrak{B}\}\\
&  =\sup\{\lambda\in\mathbb{R}\ |\ I-F_{\lambda}^{A_{\varepsilon}%
}=I-F_{\lambda_{k-1}}^{A}\}\\
&  =a_{k}.
\end{align*}
Moreover,%
\[
\mathfrak{B}\in N_{k}\Longrightarrow g_{A}(\mathfrak{B})=\sup\{\lambda
\in\mathbb{R}\ |\ I-F_{\lambda}^{A}\in\mathfrak{B}\}\in\lbrack\lambda
_{k-1},\lambda_{k}].
\]
For all $\mathfrak{B}\in\mathcal{Q(R)}$, we thus have $|(g_{A_{\varepsilon}%
}-g_{A})(\mathfrak{B})|<\varepsilon$, i.e.%
\[
|g_{A_{\varepsilon}}-g_{A}|_{\infty}\leq\varepsilon,
\]
so the antonymous function $g_{A}$ of $A\in\mathcal{R}_{sa}$ is continuous.
\end{proof}

\ 

Let $\mathcal{A(R)}:=\{g_{A}\ |\ A\in\mathcal{R}_{sa}\}$ denote the set of
antonymous functions of $\mathcal{R}$, and let $\mathcal{O(R)}:=\{f_{A}%
\ |\ A\in\mathcal{R}_{sa}\}$ denote the set of observable functions of
$\mathcal{R}$. We have shown that the set $\mathcal{A(R)}$ of antonymous
functions is a subset of $C_{b}(\mathcal{Q(R)},\mathbb{R})$, the bounded
continuous real-valued functions on the Stone spectrum $\mathcal{Q(R)}$ of
$\mathcal{R}$. $\mathcal{A(R)}$ separates the points of $\mathcal{Q(R)}$,
since $g_{P}=\chi_{\mathcal{Q}_{P}(\mathcal{R})}$ for a projection
$P\in\mathcal{P(R)}$. $\mathcal{A(R)}$ contains the constant functions (choose
$A=0$ in Lemma \ref{Lg_(A+aI)=a+g_A}). Since $g_{A}=1-f_{I-A}=-f_{-A}$ (see
section \ref{_SMotivation}), we have $\mathcal{A(R)}=-\mathcal{O(R)}$.

\subsection{Antonymous functions as generalized Gelfand transforms}

Let $\mathcal{R}$ be an abelian von Neumann algebra. Thm. 2.9 in \cite{deG05b}
shows that the mapping
\begin{align*}
\omega:\mathcal{R}_{sa}  &  \longrightarrow\mathcal{O(R)}\\
A  &  \longmapsto f_{A},
\end{align*}
sending a self-adjoint operator $A$ to its observable function $f_{A}$, is the
restriction of the Gelfand transformation to the self-adjoint operators in
$\mathcal{R}$. Here, the Gelfand spectrum $\Omega(\mathcal{R})$ and the Stone
spectrum $\mathcal{Q(R)}$ of $\mathcal{R}$ are identified using thm. 3.2 of
\cite{deG05}. Let%
\[
\theta:\mathcal{Q(R)}\longrightarrow\Omega(\mathcal{R})
\]
denote the homeomorphism between the Stone spectrum and the Gelfand spectrum
of $\mathcal{R}$. Since the Gelfand spectrum $\Omega(\mathcal{R})$ is compact,
so is the Stone spectrum $\mathcal{Q(R)}$, and all continuous functions on
$\Omega(\mathcal{R})$ (resp. $\mathcal{Q(R)}$) are bounded.

\ 

Let $\Gamma:\mathcal{R}\rightarrow C(\Omega(\mathcal{R}))$ denote the Gelfand
transformation. This is an isometric $\ast$-isomorphism, so in particular,
$\Gamma(\mathcal{R}_{sa})=C(\Omega(\mathcal{R}),\mathbb{R})$, i.e. the
self-adjoint operators in $\mathcal{R}$ are mapped bijectively to the
real-valued continuous functions on the Gelfand spectrum $\Omega(\mathcal{R}%
)$. The homeomorphism $\theta$ from $\Omega(\mathcal{R})$ onto $\mathcal{Q(R)}%
$ induces a $\ast$-isomorphism
\begin{align*}
\theta^{\ast}:C(\Omega(\mathcal{R}))  &  \longrightarrow C(\mathcal{Q(R)})\\
h  &  \longmapsto h\circ\theta.
\end{align*}

Prop. \ref{PAbVNAg_A=f_A} shows that for all $A\in\mathcal{R}_{sa}$, the
antonymous function $g_{A}$ coincides with the observable function $f_{A}$ if
$\mathcal{R}$ is abelian. Hence we get:

\begin{theorem}
\label{TAbVNAOFsAreGelfTransfsOfSelfAdjOps}Let $\mathcal{R}$ be an abelian von
Neumann algebra. Then the mapping%
\begin{align*}
\alpha:\mathcal{R}_{sa}  &  \longrightarrow\mathcal{A(R)}\subseteq
C(\mathcal{Q(R)},\mathbb{R}),\\
A  &  \longmapsto g_{A},
\end{align*}
from $\mathcal{R}_{sa}$ to $C(\mathcal{Q}\mathcal{(R)},\mathbb{R})$, sending a
self-adjoint operator $A$ to its antonymous function $g_{A}$, is the
restriction of the Gelfand transformation $\Gamma$ to $\mathcal{R}_{sa}$.
Here, the $\ast$-isomorphism $\theta^{\ast}$ is used to identify the
homeomorphic spaces $\Gamma(\mathcal{R}_{sa})=C(\Omega(\mathcal{R}%
),\mathbb{R})$ and $C(\mathcal{Q(R)},\mathbb{R})$.
\end{theorem}

\begin{corollary}
Let $\mathcal{R}$ be an abelian von Neumann algebra. Then the mapping
$\alpha:\mathcal{R}_{sa}\rightarrow C(\mathcal{Q(R)},\mathbb{R})$, $A\mapsto
g_{A}$, is surjective, i.e. every real-valued continuous function on the Stone
spectrum $\mathcal{Q(R)}$ is an antonymous function.
\end{corollary}

\begin{proof}
Since $\theta^{\ast}:$ $C(\Omega(\mathcal{R}))\rightarrow C(\mathcal{Q(R)})$
is a $\ast$-isomorphism,$\ $we have $C(\mathcal{Q(R)},\mathbb{R)}=\theta
^{\ast}(C(\Omega(\mathcal{R}),\mathbb{R})=\theta^{\ast}(\Gamma(\mathcal{R}%
_{sa}))$, i.e. every function $\varphi\in C(\mathcal{Q(R)},\mathbb{R})$ is of
the form $\theta^{\ast}(\widehat{A})$, where $\widehat{A}=\Gamma(A)$ is the
Gelfand transform of some self-adjoint operator $A\in\mathcal{R}_{sa}$.
\end{proof}

\begin{remark}
Of course, the mappings $\omega$ and $\alpha$ can also be considered for
arbitrary, non-abelian von Neumann algebras. Then these mappings are two
different \emph{non-commutative generalizations of the Gelfand transformation}
(for self-adjoint elements).
\end{remark}

Let $\mathcal{R}$ be an arbitrary von Neumann algebra. Since every operator
$B\in\mathcal{R}$ has a unique decomposition%
\[
B=A_{1}+iA_{2},
\]
where $A_{1},A_{2}$ are self-adjoint operators in $\mathcal{R}$, the mapping%
\begin{align*}
\alpha:\mathcal{R}_{sa} &  \longrightarrow C(\mathcal{Q(R)},\mathbb{R)}\\
A &  \longmapsto g_{A}%
\end{align*}
can be extended canonically to a mapping%
\begin{align*}
\alpha^{\prime}:\mathcal{R} &  \longrightarrow C(\mathcal{Q(R)}\mathbb{)}\\
B &  \longmapsto g_{A_{1}}+ig_{A_{2}}.
\end{align*}
If $\mathcal{R}$ is abelian, then this mapping is the Gelfand transformation
of $\mathcal{R}$ (using the identification of the Stone spectrum and the
Gelfand spectrum).

\ 

Addition and multiplication of antonymous functions can be defined pointwise,
and thus $\mathcal{A(R)}$ becomes a $\mathbb{R}$-linear space. But of course,
for a non-abelian von Neumann algebra $\mathcal{R}$, the mapping
$\alpha:\mathcal{R}_{sa}\longrightarrow\mathcal{A(R)}\subseteq
C(\mathcal{Q(R)},\mathbb{R)}$ does not respect the linear structure of
$\mathcal{R}_{sa}$. This mapping is $\mathbb{R}$-homogeneous, but we have
$g_{A_{1}+A_{2}}\neq g_{A_{1}}+g_{A_{2}}$ in general ($A_{1},A_{2}%
\in\mathcal{R}_{sa}$). Consequently, the mapping $\alpha^{\prime}%
:\mathcal{R}\longrightarrow C(\mathcal{Q(R)})$, $\alpha(B)=g_{A_{1}}%
+ig_{A_{2}}$ is \emph{not} an algebra homomorphism unless $\mathcal{R}$ is abelian.

\ 

\textbf{\textbf{\textsf{Acknowledgement}}}

\ 

The author would like to thank Martin D\"{o}ring for helpful comments and
Chris Isham, Joachim Weidmann and Dennis Dietrich for their interest.

\end{document}